\documentclass[9pt,twocolumn,twoside]{osajnl}

\journal{ol} 
\usepackage{graphicx}
\usepackage{dcolumn}
\usepackage{lipsum}
\usepackage{bm}
\usepackage{tabu}
\usepackage[utf8]{inputenc}
\usepackage[T1]{fontenc}

\usepackage{color}

\usepackage{comment}
\usepackage{soul}

\setboolean{shortarticle}{true}

\title{Controlled longitudinal spin-orbit separation of complex vector modes}

\author[1,*]{Xiao-Bo Hu}
\author[2]{Bo Zhao}
\author[1]{Rui-Pin Chen}
\author[3,*]{Carmelo Rosales-Guzm\'an}
\affil[1]{Key Laboratory of Optical Field Manipulation of Zhejiang Province,Department of Physics, Zhejiang Sci-Tech University, Hangzhou, 310018, China}
\affil[2]{Wang Da-Heng Collaborative Innovation Center for Quantum manipulation \& Control, Harbin University of Science and Technology, Harbin 150080, China}
\affil[3]{Centro de Investigaciones en Óptica, A.C., Loma del Bosque 115, Colonia Lomas del campestre, 37150 León, Gto., Mexico}
\affil[*]{Corresponding author:huxiaobo@zstu.edu.cn, carmelorosalesg@hrbust.edu.cn}

\begin{abstract} 
Complex vector modes, entangled in spin and orbital angular momentum, are opening burgeoning opportunities for a wide variety of applications. Importantly, the flexible manipulation the various properties of such beams will pave the way to novel applications. As such, in this manuscript, we demonstrate a longitudinal spin-orbit separation of complex vector modes propagating in free space. To achieve this we employed the recently demonstrated circular Airy Gaussian vortex vector (CAGVV) modes, which feature a self-focusing property. More precisely, by properly manipulating the intrinsic parameters of CAGVV modes, the strong coupling between the two constituting orthogonal components of CAGVV mode undergo a spin-orbit separation along the propagation direction namely, while one polarisation component, focuses at a specific plane, the other focuses at a different plane. Such spin-orbit separation, which we demonstrated by numerical simulations and corroborated experimentally, can be adjusted on-demand by simply changing the initial parameters of CAGVV modes. Our findings will be of great relevance, for example in optical tweezers, to manipulate micro- or nano-particles at two different parallel planes. 
\end{abstract}

\setboolean{displaycopyright}{true}

\begin{document}
\maketitle

The field of structured light, which encompasses the manipulation of the various properties of light, resulting into novel light fields, has quickly became an important branch of modern optics  \cite{Roadmap,Shen2022}. In particular, there is an increasing interest in generating non-separable states of light, using the spatial and polarisation degrees of freedom \cite{Rosales2018Review,rosales2021Mathieu,Zhao2022}. One of the reasons is the wide variety of applications they have found in various areas of research such as, optical tweezers, optical communications, optical metrology, high resolution microscopy, amongst many others \cite{Zhan2009,hu2019situ,li2016high,Yuanjietweezers2021,Ndagano2017,bhebhe2018,zhao2015high,Ndagano2018}. Another reason is the high similarity they share with entangled states, which has earned them the controversial term "classically-entangled" modes \cite{Spreeuw1998,Karimi2015,toninelli2019,konrad2019,ChavezCerda2007}. Such similarity has also motivated several quantum-inspired applications \cite{Toppel2014,Shen2022,BergJohansen2015,forbes2019classically}. In particular, in the last years it has become topical the study of vector modes with a varying polarisation states along the propagation direction, which provides with a potential tool for novel applications  \cite{Moreno2015,ShiyaoFu2016,PengLi2017,PengLi2016}. Here for example, vector modes that oscillate from one vector state to another, while keeping a constant degree of non-separability have been reported, as well as the case where the degree of non-separability oscillates between scalar and vector, paving the path to the on-demand delivery of specific vector states \cite{Otte2018}. In some other cases, not only the polarisation distributions changes, but also the topological charge of the vector mode \cite{Davis2016}. Additional examples include the case where the polarisation distribution of the vector field oscillates as function of the propagation distance relaying in the Gouy phase or in the use of frozen waves \cite{zhong2021,PengLi2018}. A more recent example reported the case of a pure vector beam with parabolic shape that upon propagation split into two spatially-disjoint modes of orthogonal circular polarisation\cite{Hu2021}. In this particular work, the separation of the modes happens in the transverse plane, resembling the spin-Hall effect. 

In parallel to the above, abruptly autofocusing Circular Airy Gaussian vortex (CAGV) beams have also been the subject of several works since their inception in 2010 \cite{Efremidis2010}. Such modes, which result as the superposition of a large number of airy beams arranged in a circular shape, have the ability to self-focus upon propagation, following a parabolic trajectory  \cite{Chen2015}. Generally, these modes are generated digitally using spatial light modulators for a total control of their propagation dynamics. For example, the beam's diameter in the focal plane can be tuned as function of the order of the topological charge. In addition, the focusing distance can be controlled by modifying the launch angle,  which is directly related to the ballistic propagation trajectory of the beam. Additionally, in recent time, the vectorial version of CAGV modes was demonstrated, providing a new class of vector beams with a self-focusing property \cite{Hu2022SciRep}. 

In this manuscript we report a novel property of vector beams, the propagation-dependent separation of its two constituting orthogonal spatial modes into two scalar modes of opposite spin and orbit orbital Angular Momentum (AM), a phenomena we termed spin-orbit separation of vector modes. It is worth mentioning that a spin-based scalar version was reported before \cite{Liu2016}, nonetheless the principle governing the separation demonstrated in this manuscript is completely different. Further, properties such as the separation distances can be dynamically controlled by means of the parameters of the CAGV beams.

\begin{figure}[tb]
    \centering
    \includegraphics[width=0.47\textwidth]{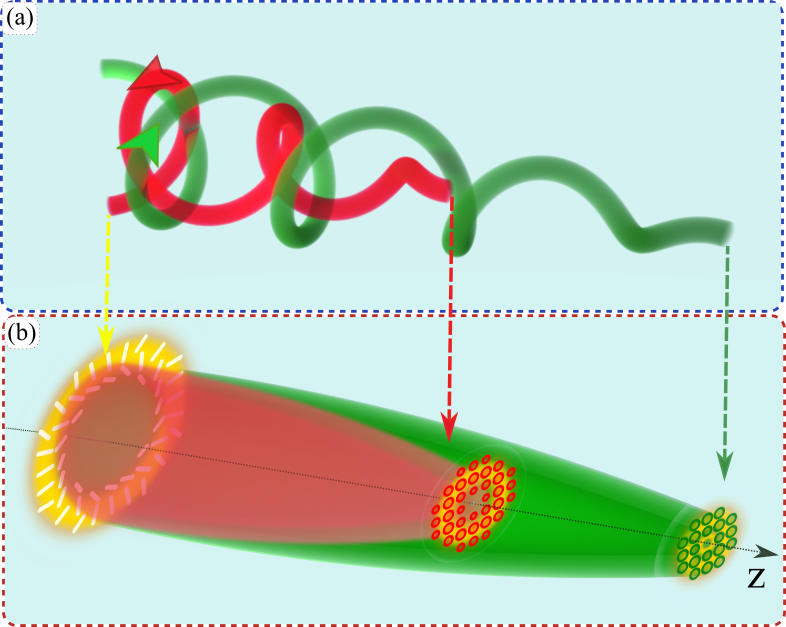}
    \caption{ Schematic representation of the longitudinal spin-orbit separation of vortex vector beams, illustrating the propagation-dependent (a) separation of spin (associated to polarisation) and (b) the beam's intensity and polarisation distribution at three different planes, featuring an auto-focusing effect of each constituting spatial mode. Here the red and green colors represent the right- and left-handed circular polarization, respectively. Further, linear polarisation is represented by white lines.}
    \label{concept}
\end{figure}

Figure \ref{concept} shows an schematic representation of the spin-orbit separation introduced in this manuscript, whereby, upon propagation, the circular Airy Gaussian vortex vector (CAGVV) beam undergoes a dynamic decoupling from a maximum degree of nonseparability to a null, featuring a clear separation of right- and left-handed circular polarization components, as schematically depicted in Fig.\ref{concept}(a). This phenomena is enabled by the self-accelerating property of CAGV beams, whereby each orthogonal spin-orbit component auto-focuses at two different parallel planes. For the sake of clarity, Fig.\ref{concept}(b), shows the polarization distribution of CAGVV beam at three different planes, at the origin, where the vector beam features only linearly polarized states(marked as white lines), and at two consecutive planes where the beam separates into its right- and left-handed circular polarization components, represented by red and green ellipses. Noteworthy the spin-orbit separation along the propagation direction can be controlled on-demand by properly setting the beam's inherent parameters, as we will explain below.

Mathematically, the CAGV beam can be defined in cylindrical coordinates as \cite{Hu2022SciRep},

\begin{equation}\label{Eq:CAV}
    {\bf U}_{m}^{\nu}({\bf r},\phi)={r^{\lvert m \rvert}}Ai\left(\frac{r_0-r}{\omega} \right)\exp\left({\frac{a(r_0-r)}{\omega}}\right){\exp(im\phi)}\exp({i\nu r}),
\end{equation}
where $Ai({\cdot})$ is the Airy function, $\omega$ is a scaling factor, $a$ is a truncation parameter and $r_0$ is the beam radius in the initial plane. Another important parameter is the initial launch angle, denoted with $\nu$, which plays a key role in the self-focusing effect when the beam propagates. To be more specific, for $\nu<0$ the CAGV beam experiences a defocusing behaviour, while for  $\nu>0$ a focusing behaviour. Additionally, such CAGV beam features an azimuthally-varying phase of the form $\exp({im\phi})$, where the index $m$ is known as the topological charge, associated to an $m\hbar$ amount of Orbital Angular Momentum (OAM) per photon, with $\hbar$ being the reduced Plank constant. 

Following the general procedure to generate complex vector modes as a non-separable weighted superposition of the spatial and polarization degrees of freedom(DoF)\cite{Galvez2012,Galvez2015}, we generate CAGVV modes by encoding the CAGV beams with different values of OAM as the spatial DoF and one set of orthogonal and opposite spin as the polarization DoF. Mathematically, such superposition takes the forms, 
\begin{equation}
    \centering
    {\bf U}_{m_{1,2}}^{\nu_{1,2}}({\bf r},\phi)=\cos\theta {\bf U}_{m_1}^{\nu_1}({\bf r},\phi)\hat{\bf e}_R+\sin\theta\exp(i\alpha) {\bf U}_{m_2}^{\nu_2}({\bf r},\phi)\hat{\bf e}_L,
    \label{Eq:VM}
\end{equation}
where $\hat{\bf e}_R$ and $\hat{\bf e}_L$ are unitary vectors representing the right and left circular states of polarization, respectively. The ${\bf U}_{m_1}^{\nu_1}({\bf r},\phi)$ and ${\bf U}_{m_2}^{\nu_2}({\bf r},\phi)$, represent the spatial degree of freedom, which are weighted by $\theta\in[0,\pi/2]$. Both modes carry a fix amount of OAM, $m_1\hbar$ and $m_2\hbar$, respectively. Particularly, the trajectory of each can also be manipulated flexibly by the initial launch angles, $\nu_1$ and $\nu_2$. An additional inter-modal phase $\exp(i\alpha)$, with $\alpha\in [0, \pi]$, sets a phase delay between both polarisation components. From now on, we will omit the explicit dependence of $({\bf r},\phi)$, unless is necessary.
\begin{figure}[tb]
    \centering
    \includegraphics[width=0.48\textwidth]{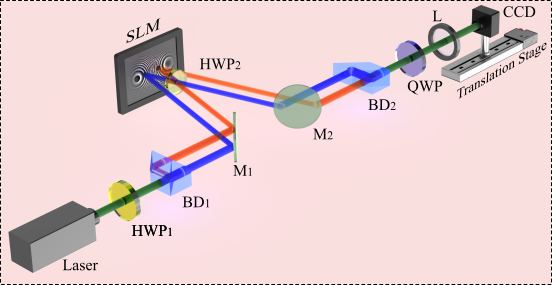}
    \caption{Schematic representation of the optical setup implemented to generate CAGVV mode.  HWP$_1$-HWP$_2$: Half-Wave plate, BD$_1$-BD$_2$: Beam displacer, M$_1$-M$_2$: Mirror, QWP: Quarter-Wave Plate, SLM: Spatial Light Modulator, L: lenses, CCD: Charged Device Camera.
}
    \label{setup} 
\end{figure}

To generate experimentally CAGVV beams we implemented the optical setup depicted in Fig.\ref{setup}. Here, an expanded and collimated laser beam  ($\lambda=532$nm) with horizontal linear polarisation is first rotated to diagonal polarisation with the use of a half-wave plate (HWP$_1$) at $22.5^\circ$. A beam displacer (BD$_1$) subsequently separates the beam into its horizontal (blue color) and vertical (orange color) polarization components, afterwards both are directed to a spatial light modulator (SLM) using a mirror (M$_1$). Here, the screen of the SLM is digitally split into two horizontal sections, each of which is addressed with independent holograms that encodes the Fourier transform of the constituting scalar fields ${\bf U}_{m_1}^{\nu_1}$ and ${\bf U}_{m_2}^{\nu_2}$. Since the SLM only modulates the horizontal polarization, a second half-wave plate (HWP$_2$) is placed before the SLM in the path of the vertically polarized beam to rotate its polarisation to horizontal. Meanwhile, the other beam, which is first modulated by the SLM, then is sent through the same HWP$_2$ to transform its polarization into vertical. Finally, a second mirror (M$_2$) redirects both beams to another beam displacer (BD$_2$), which recombines them into the desired CAGVV beam. Finally, we add a quarter-wave plate (QWP) at 45$^\circ$ to transform the CAGVV beam from the linear ($\hat{\bf e}_H$, $\hat{\bf e}_V$) to the circular ($\hat{\bf e}_R$, $\hat{\bf e}_L$) polarization basis. The intensity of the generated beams is recorded with a charged-coupled device (CCD: FL3-U3-120S3C-C with a resolution of 1.55$\mu$m), which is placed in the focal plane of L$_3$($f=300$mm) and  mounted on a rail to record the beam upon propagation.

\begin{figure}[tb]
    \centering
    \includegraphics[width=0.47\textwidth]{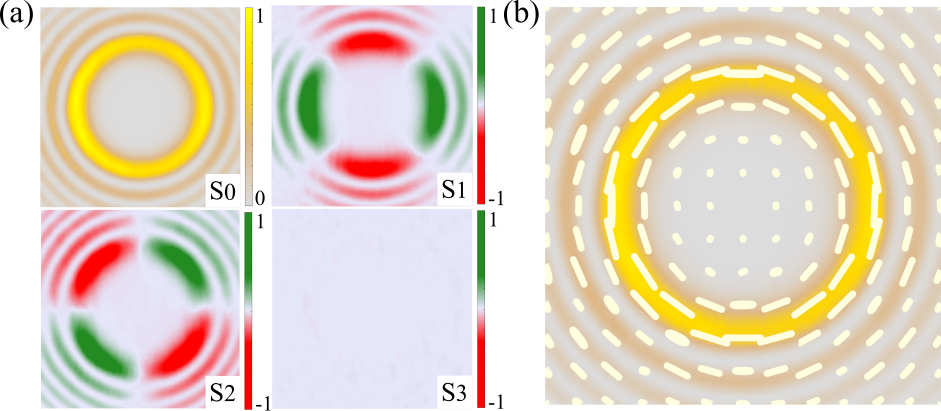}
    \caption{Experimental Stokes parameters $S_{0}$, $S_{1}$, $S_{2}$ and $S_{3}$ of the vector mode ${\bf U}_{-1, 1}^{0, 0}$ are represented in (a), where the reconstruction of polarization distribution are overlapped with the total intensity $S_{0}$ shown in (b). Here, the white color presents the horizontal ellipses. For this example, $\omega$=0.2mm, $r_0$=1mm, $a$=0.4.}
    \label{Stokesrecons} 
\end{figure}

To conduct a detailed analysis of the propagation dynamics of the generated CAGVV beams, we reconstructed the polarisation distribution of the beam using Stokes polarimetry. To this end, we computed the Stokes parameters from intensity measurements using the relations,
\begin{equation}\label{Eq.SimplyStokes}
\begin{split}
\centering
 &S_{0}=I_{H}+I_{V},\hspace{11.5mm} S_{1}=2I_{H}-S_{0},\hspace{1mm}\\
 &S_{2}=2I_{D}-S_{0},\hspace{10mm} S_{3}=2I_{R}-S_{0}.
\end{split}
\end{equation}
Here, $I_{H}$, $I_{D}$, $I_{V}$and $I_{R}$ are the horizontal, diagonal, vertical, and right-circular, respectively polarization components of the beam. Such intensities can be experimentally measured through a series of phase retarders inserted in front of the CCD. More specifically, we measured the $I_{H}$, $I_{D}$ and $I_{V}$ by passing the generated beam through a linear polarizer originated at $0^\circ$, $45^\circ$and $90^\circ$, respectively, while $I_{R}$ can be acquired by inserting another QWP at $45^\circ$ with the polarizer setting at $90^\circ$ (see \cite{Zhao2019} for more details). As way of example, in Fig.\ref{Stokesrecons}(a) we show the experimental Stokes parameters of the vector mode ${\bf U}_{-1, 1}^{0, 0}$ with parameters $\omega$=0.2mm, $r_0$=1mm, $a$=0.4, from which the azimuthal polarization distribution can be reconstructed, as shown in Fig.\ref{Stokesrecons}(b), where the intensity distribution is also shown for reference.

\begin{figure}[tb]
    \centering
    \includegraphics[width=0.48\textwidth]{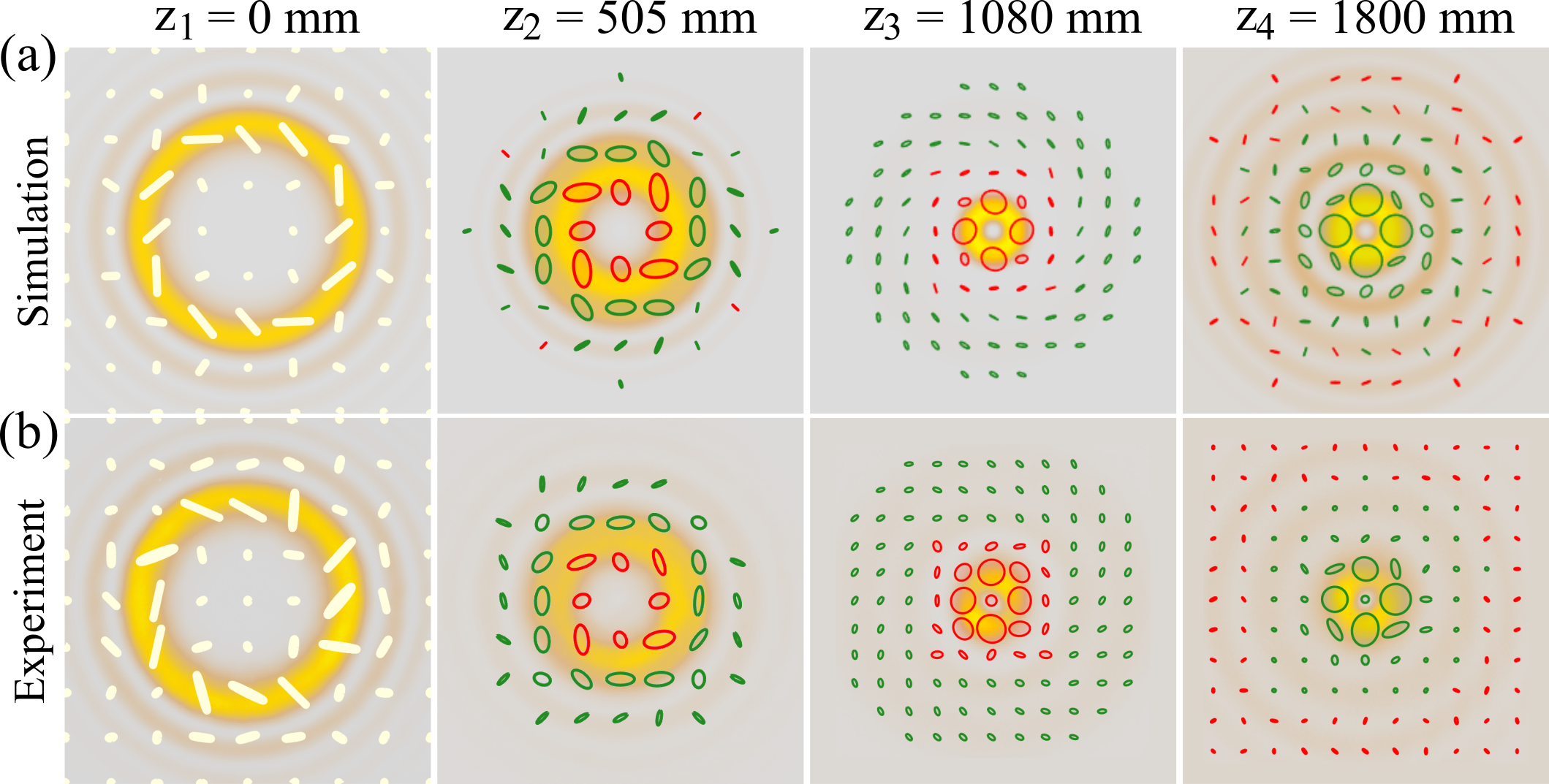}
    \caption{Polarization distribution of ${\bf U}_{1, -1}^{2, 0.5}$ mode upon free space propagation. Theoretical (a) and experimental (b) transverse intensity profiles overlapped with the polarization distribution is shown at four different planes, namely $z_1=0$mm, $z_2=505$mm, $z_3=1080$mm, $z_4=1800$mm, where the right- and left- handed polarization presented with red and green ellipses,  white lines symbolized linear polarisation. }
    \label{Stokespropagation} 
\end{figure}

To observe the longitudinal spin-orbit separation of our generated CAGVV mode, we scanned its transverse polarization distribution upon propagation. By way of example and using Stokes polarimetry, we reconstructed the transverse polarization distribution of the specific mode ${\bf U}_{1, -1}^{2, 0.5}$ at four different planes, as shown  in Fig.\ref{Stokespropagation}, numerical simulations shown in Fig.\ref{Stokespropagation}(a) and experimental results in Fig.\ref{Stokespropagation}(b). As can be seen, first in the initial plane ($z_1=0$mm), our generated vector mode features a mix of linear polarization (marked as white lines), see the left panel of Fig.\ref{Stokespropagation}(a). At the propagation distance of $z_2=505$mm, we show another example where the transverse polarization distribution features a mix of right- and left-handed elliptical polarization. As the beam further propagates, the right circular polarization component embedded with a topological charge $m_1=1$ first focuses at the transverse plane $z_3=1080$mm, as shown in the third panel, where as the left component of topological charge of $m_2=-1$, then focuses at the plane $z_4=1800$mm, as shown on the right panel. Here, right and left handed elliptical polarization components are represented with red and green color, respectively. 

\begin{figure}[tb]
    \centering
    \includegraphics[width=0.47\textwidth]{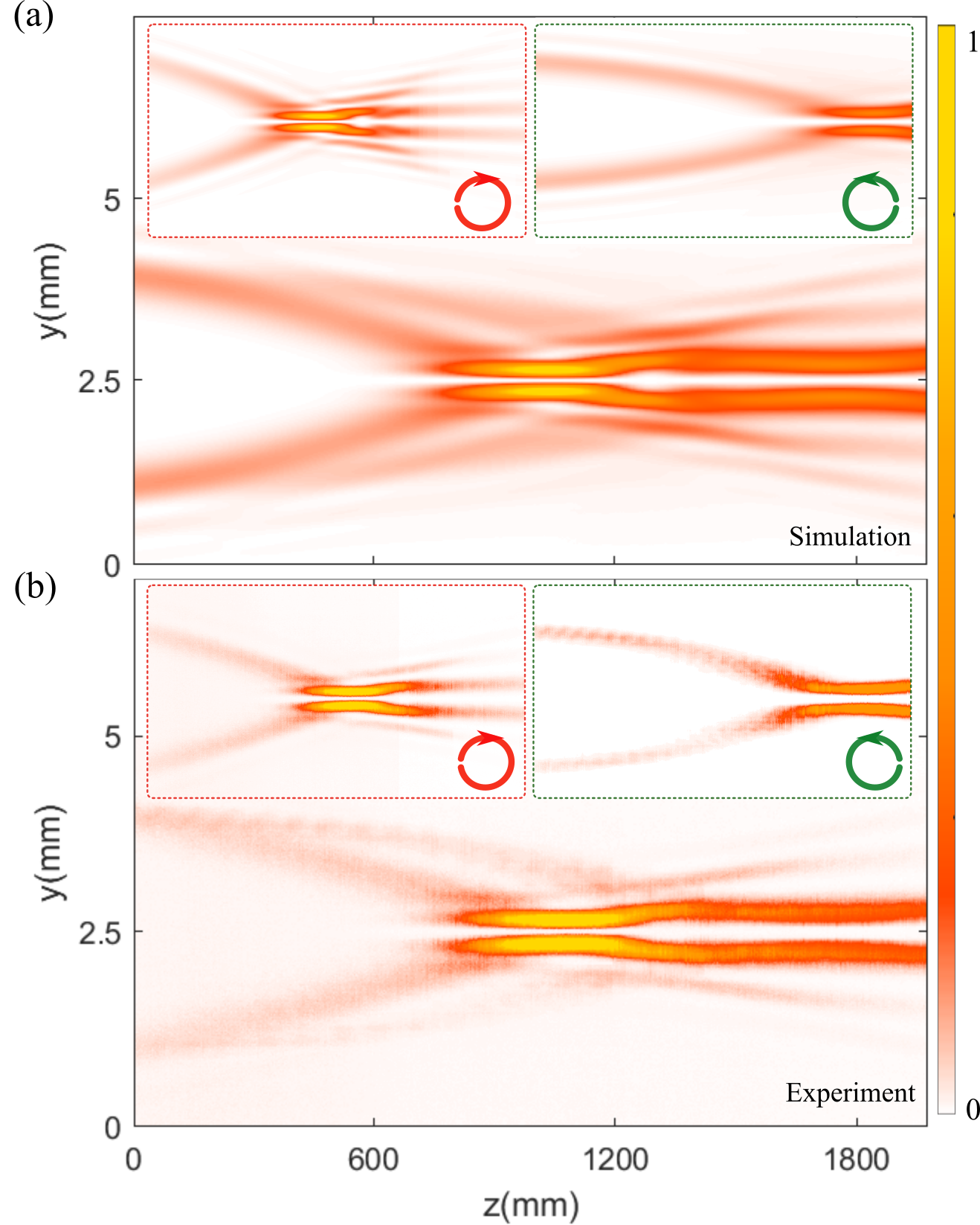}
    \caption{The propagation trajectory of the maximum intensity for the specific case of ${\bf U}_{1, -1}^{2, 0.5}$ is depicted, where two obvious auto-focus process following parabolic trajectory is observed.
    Here, the theoretical and experimental results are shown in (a) and (b), respectively. On the top of each, we also show the intensity corresponding to the right- and left-handed circular polarization components as the insets on the left and right panels,  represented as red and green circles, respectively. Here, we show the specific case for which, $\omega$=0.2mm, $r_0$=1mm, $a$=0.4.}
    \label{propagationresults} 
\end{figure}

Figure \ref{propagationresults} shows the peak intensity of the engineered ${\bf U}_{1, -1}^{2, 0.5}$ as a function of the propagation distance $z$, numerical simulations in Fig. \ref{propagationresults}(a) and experimental results in Fig \ref{propagationresults}(b). In each case, the bottom panel shows the full intensity distribution of the vector beam, where two focusing section can be observed. For the sake of clarity, on the top panels we show the intensity corresponding to each polarization component, where these two sections can be clearly observed. Again right and left handed polarisation components are represented by red and green circles, respectively. Noteworthy, both focus positions can be on-demand manipulated by carefully picking the initial launch angles for each component, namely $\nu_{1}, \nu_{2}$. In the specific case of $\nu_{1}=\nu_{2}$, the generated vector beam will focus at one single plane. Given the fact that the increasing of parameter $\nu$ will enhance the abruptly autofocusing property of CAGV mode (as shown in the insets), the difference between $\nu_{1}$ and $\nu_{2}$ of the generated CAGVV mode enables a tunable longitudinal spin-orbit separation, that is, while one polarization component carrying a fix amount of OAM with bigger initial launch angle first focus at one specific plane, the other carrying different amount of OAM then focus at a further plane (affected by the smaller initial launch angle).  Additionally, another parameter of CAGV mode, the topological charge $m$, is demonstrated to influence the focal intensity contrast\cite{Jiang:18}, which pave the way to future work on flexible manipulate each focal intensity along the spin-orbital separation of CAGVV mode.


To summarize, we introduced the concept of longitudinal spin-orbit, a kind of photonic spin-Hall effect. To acchieve this, we employed circular Airy vortex vector (CAVV) beams, generated as a weighted superposition of circular Airy Gaussian vortex (CAGV) modes and orthogonal circular polarization states. The spin-orbit longitudinal separation is enabled by the autofocusing property of CAGV beam, which can be controlled in a flexible way by means of the intrinsic parameters of the beam. This new type of beams will pave the path to novel applications, of relevance in a wide variety of fields, for example, in optical tweezers they can be used to trap micro- or nano-particles at two different planes, in laser arterial processing they can for example perform material processing also at two different planes. 
  
\section*{Funding}
This research was supported by Zhejiang Provincial Natural Science Foundation of China under Grant No. LQ23A040012, Science Foundation of Zhejiang Sci-Tech University (ZSTU) under Grant No. 22062025-Y and the National Natural Science Foundation of China (61975047).

\section*{Disclosures}
The authors declare that there are no conflicts of interest related to this article.


\end{document}